\documentclass[a4paper]{jpconf}
\usepackage{graphicx}
\usepackage{cite}
\usepackage{subfigure}
\begin{document}
\title{Measurements of Heavy Cosmic-Ray Nuclei Spectra with CALET on the ISS}

\author{Yosui Akaike for the CALET Collaboration}

\address{CRESST/NASA Goddard Space Flight Center and University of Maryland Baltimore County, 8800 Greenbelt Road, Greenbelt, MD 20770, USA}

\ead{yosui.akaike@nasa.gov}

\begin{abstract}
CALorimetric Electron Telescope (CALET) has been accumulating data of high-energy cosmic rays on the International Space Station since August 2015. In addition to the primary observation of the all-electron spectra, CALET also measures the spectra of nuclei, their relative abundances and secondary-to-primary ratios to the highest energy region ever directly observed in order to investigate details of their origin and propagation in the galaxy.
The CALET instrument consists of two layers of segmented plastic scintillators to identify the individual elements from $Z=1$ to 40, a 3 radiation length thick tungsten-scintillating fiber imaging calorimeter to obtain complementary charge and tracking information, and a 27 radiation length thick segmented PWO calorimeter to measure the energy. In this paper, the capability of CALET to perform nuclei measurements and preliminary energy spectra of heavy nuclei components using 962 days of data is presented.
\end{abstract}

\section{Introduction}
Precision measurements of cosmic-ray spectra and their flux ratios are important to understand their origin and propagation in our galaxy. While the acceleration via supernova shock waves and diffusive propagation in the galactic magnetic fields are commonly accepted as a plausible scenario for cosmic-ray origins, there are many outstanding questions about the details.
In particular, recent observations such as CREAM~\cite{cream1,cream2,cream3}, PAMELA~\cite{pamela1} and AMS-02~\cite{ams1,ams2,ams3,ams4} reported that a deviation from a single power-law has been observed for proton, helium and light nuclei at a few hundred GeV/n which indicates an unexpected hardening of the spectra.
\par
CALET is a cosmic ray experiment on the International Space Station~\cite{calet1, calet2}, and has been collecting science data since mid-October 2015~\cite{calet5}.
The detector is optimized to measure electron spectrum in the multi-TeV region and the results of electron spectrum measurement have been reported~\cite{calet3, calet4}.
CALET also measures the energy spectra and elemental composition of cosmic-ray nuclei from proton to iron in the range from a few tens of GeV to the PeV scale.
The features of the CALET instrument include very good energy resolution provided by its thick calorimeter and excellent charge resolution and robust track identification based on the segmented scintillator paddles and scintillating fibers.
Also the dynamic range of CALET covers 6 orders of magnitude in energy range from 1~MIP to 1~PeV shower energy.
In this paper, we present the analysis procedure of nuclei measurements and preliminary energy spectra of heavy nuclei components with the data obtained in the period from October 13, 2015 to May 31, 2018.

\section{CALET instrument}
\begin{figure}[htbp]
  \begin{minipage}{17pc}
    \begin{center}
      \includegraphics[width=19pc]{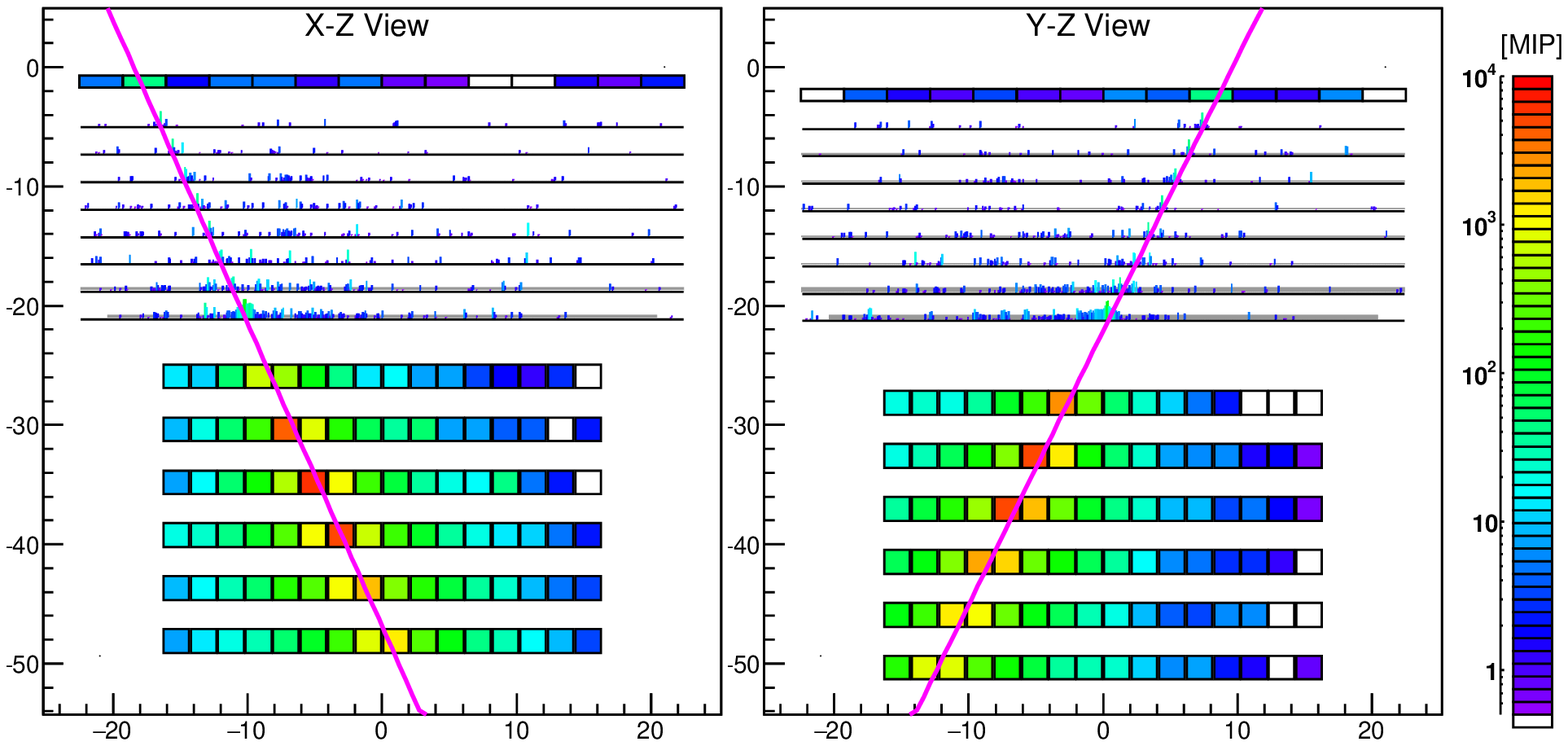}
    \end{center}
    \caption{\label{eve1}An example of carbon event with $\Delta E_{\rm{TASC}} = 1.4~\rm{TeV}$.}
  \end{minipage}\hspace{2pc}%
  \begin{minipage}{17pc}
    \begin{center}
      \includegraphics[width=19pc]{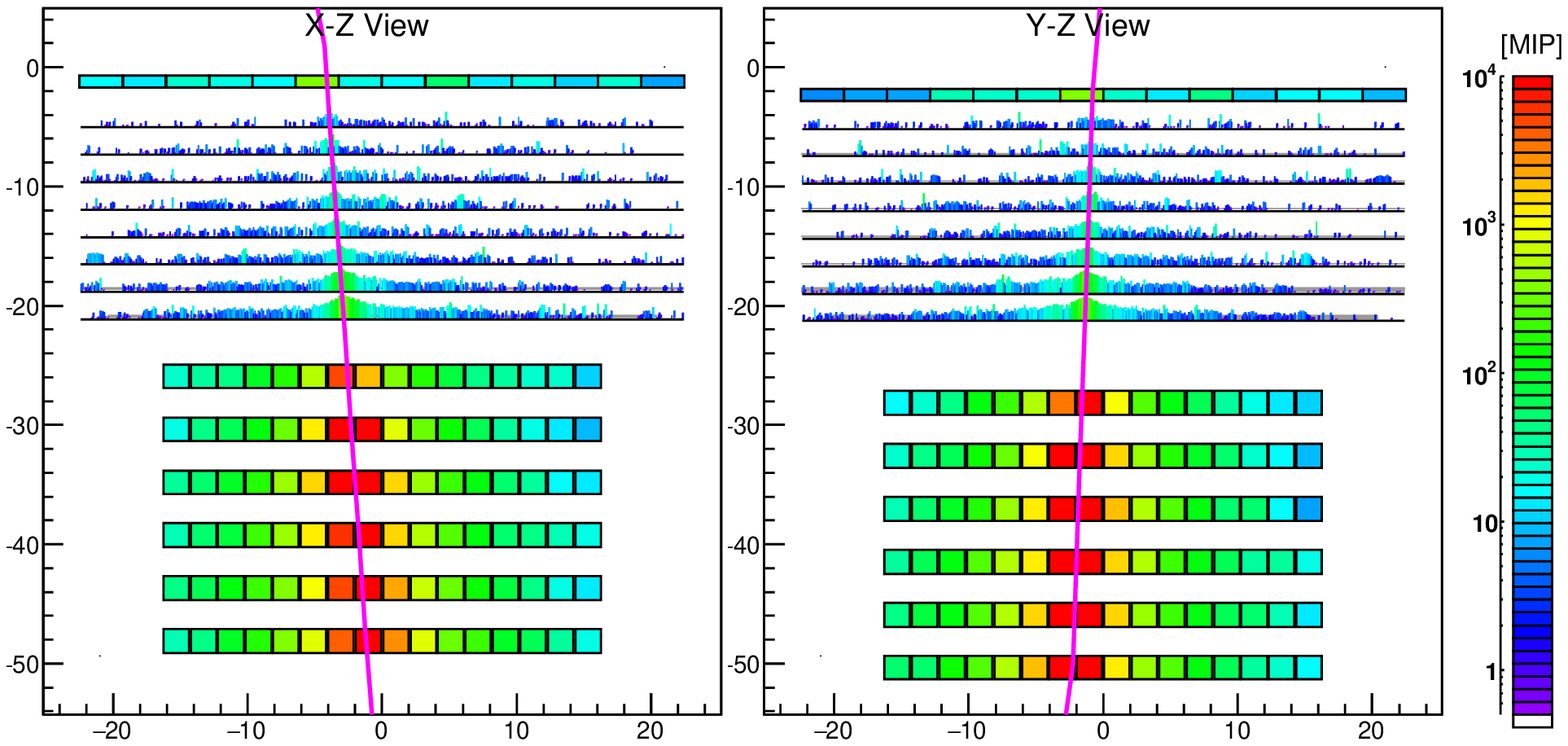}
    \end{center}
    \caption{\label{eve2}An example of iron event with $\Delta E_{\rm{TASC}} = 9.2~\rm{TeV}$.}
  \end{minipage} 
\end{figure}

The CALET detector consists of three subsystems; CHarge Detector (CHD), IMaging Calorimeter (IMC) and Total AbSorption Calorimeter (TASC).
CHD located at the top part is composed of two layers of 14 plastic scintillator paddles for measurement of the primary particle charge.
Each scintillator has dimensions of 32 mm $\times$ 450 mm $\times$ 10 mm.
IMC is a sampling calorimeter composed of ($X$, $Y$) $\times$ 8 layers of scintillating fiber (SciFi) belts and 7 tungsten plates.
Each layer of SciFi belts is made of 448 SciFis with a 1~mm square cross section and 448~mm in length.
The tungsten plates interleaved between the SciFi layers have thickness of 0.2 $X_0$ $\times$ 5 layers and 1.0 $X_0$ $\times$ 2 layers from top to bottom.
The IMC is used for track reconstruction and charge identification.
TASC is a total absorption calorimeter made of 12 layers of PWO scintillator logs for energy measurement and discrimination of electromagnetic shower and hadronic shower. Each layer has 16 PWO logs and each log has dimensions of 19 mm $\times$ 326 mm $\times$ 20 mm.
The total thickness of the calorimeter is 30~$X_0$ for electromagnetic particles or 1.3~$\lambda_{I}$ for protons.
Figure~\ref{eve1} and~\ref{eve2} show examples of schematic view of the detector with shower images of carbon and iron nuclei from the flight data.

\section{Data analysis}
\begin{figure}
  \hspace{2pc}
  \begin{center}
    \subfigure[Carbon]{ \includegraphics[width=18pc]{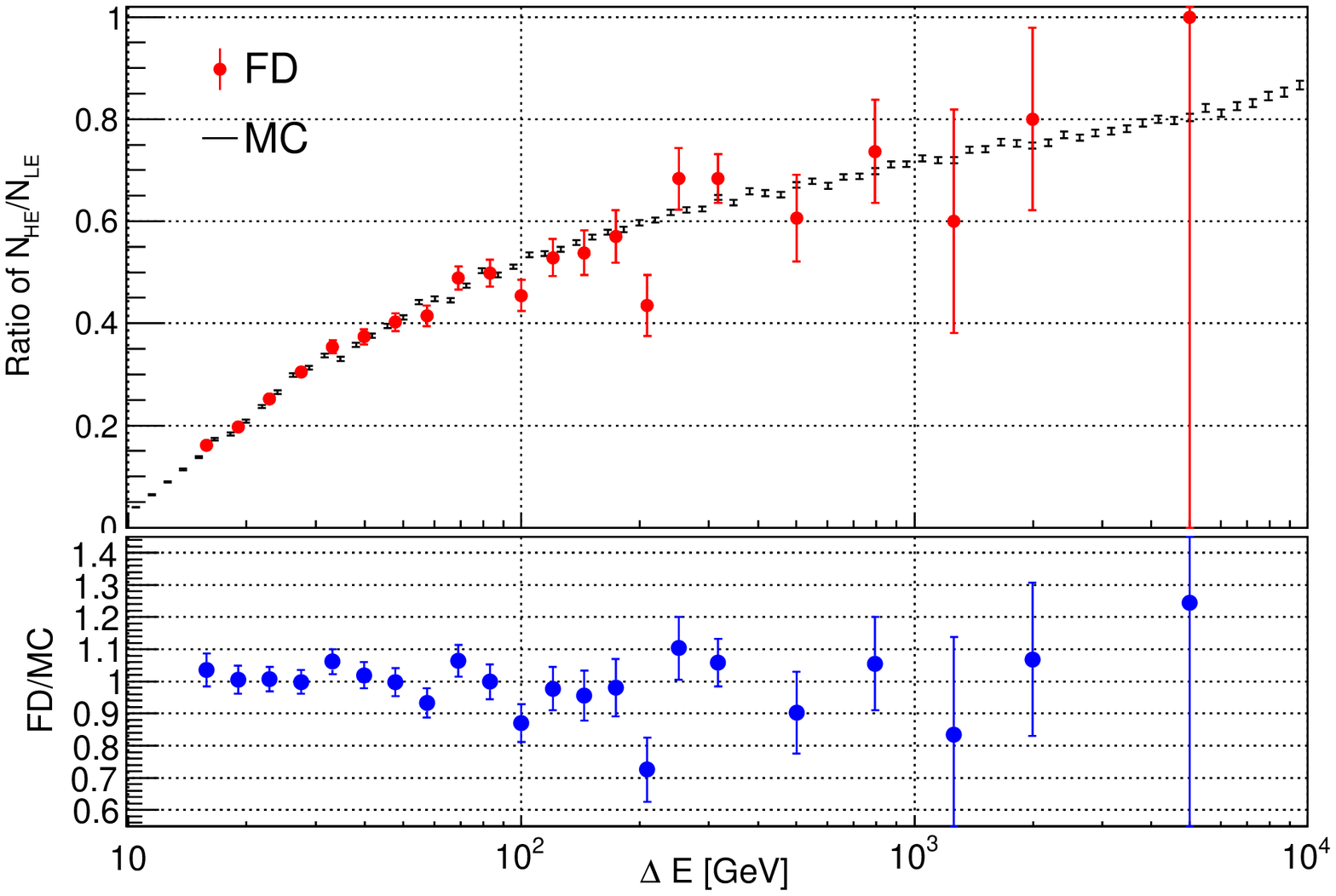}}
    \subfigure[Oxygen]{ \includegraphics[width=18pc]{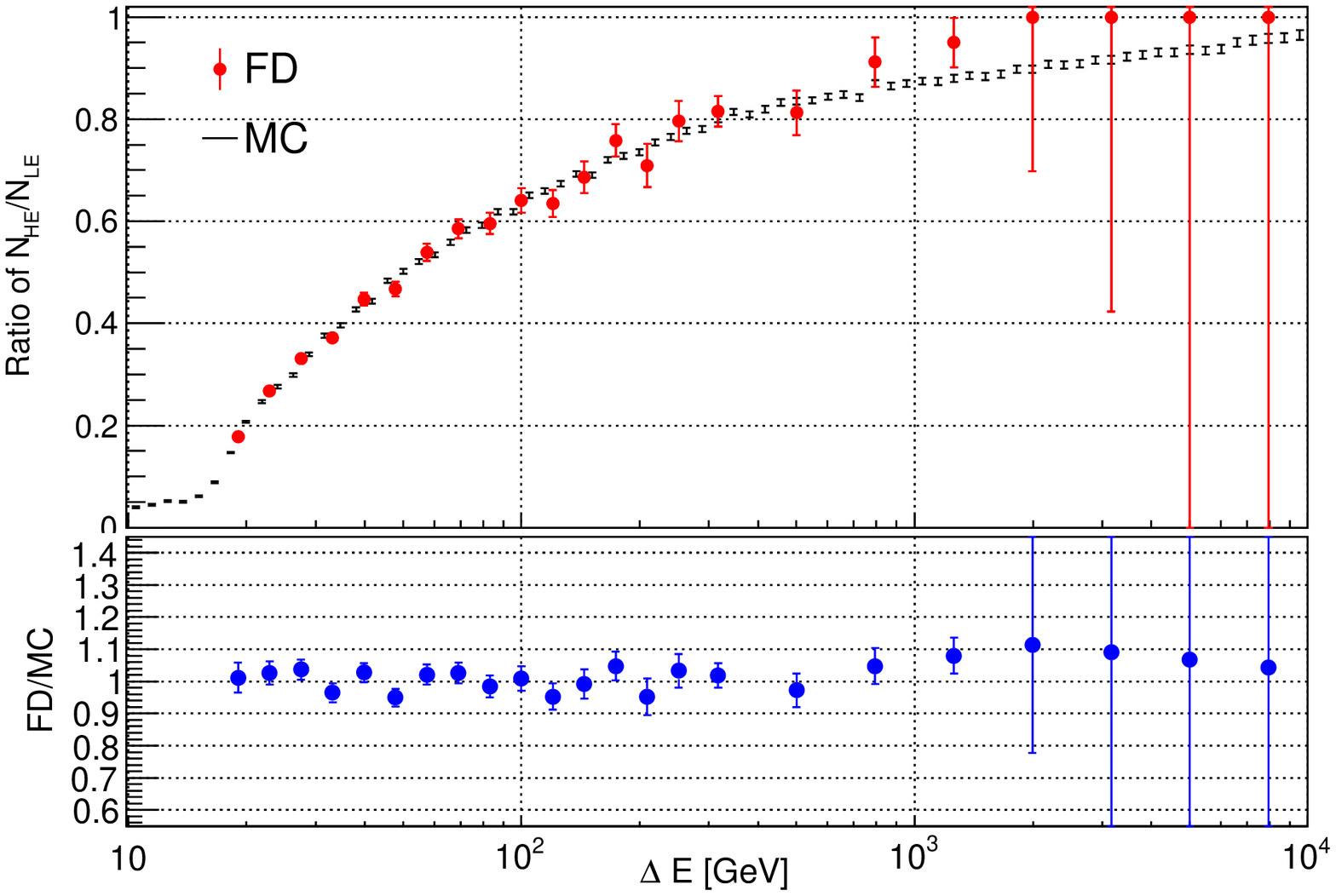}}
  \end{center}
  \caption{High energy trigger efficiency of flight data (FD) compared with MC simulations.}
  \label{trigeff}
\end{figure}

\paragraph{Event trigger}
The events detected by the onboard high energy trigger (HET) are used for the flux analysis.
HET requires a coincidence of the two bottom layers of IMC and the top layer of TASC with thresholds to detect electrons above 10 GeV.
Therefore, for light nuclei only events with a shower induced in the detector are triggered by HET, while for heavier nuclei with $Z \ge 10$ also penetrating particles are triggered because their $dE/dx$ is large enough to exceed the trigger threshold.
HET efficiency for light nuclei is evaluated using a data set taken with the same trigger logic, but with a lower threshold that allows to detect penetrating particles.
Figure~\ref{trigeff} shows the HET efficiency compared with MC simulation. Data and MC are in good agreement.

\begin{figure}[htbp]
  \begin{minipage}{16pc}
    \begin{center}
      \includegraphics[width=16pc]{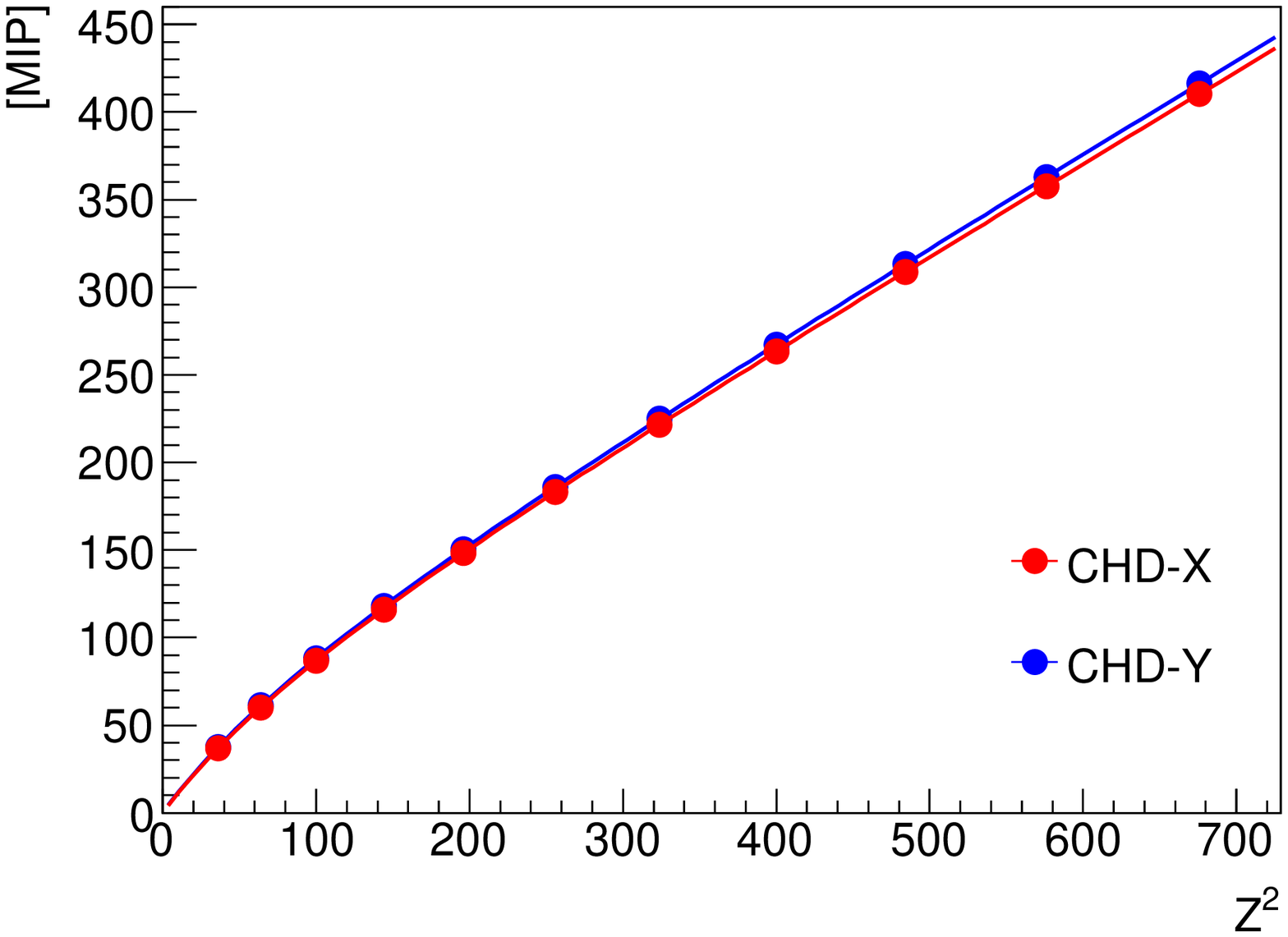}
    \end{center}
    \caption{\label{quench}Peak value of CHD signals of some charge as a function of $Z^2$.}
  \end{minipage}\hspace{2pc}%
  \begin{minipage}{20pc}
    \begin{center}
      \includegraphics[width=20pc]{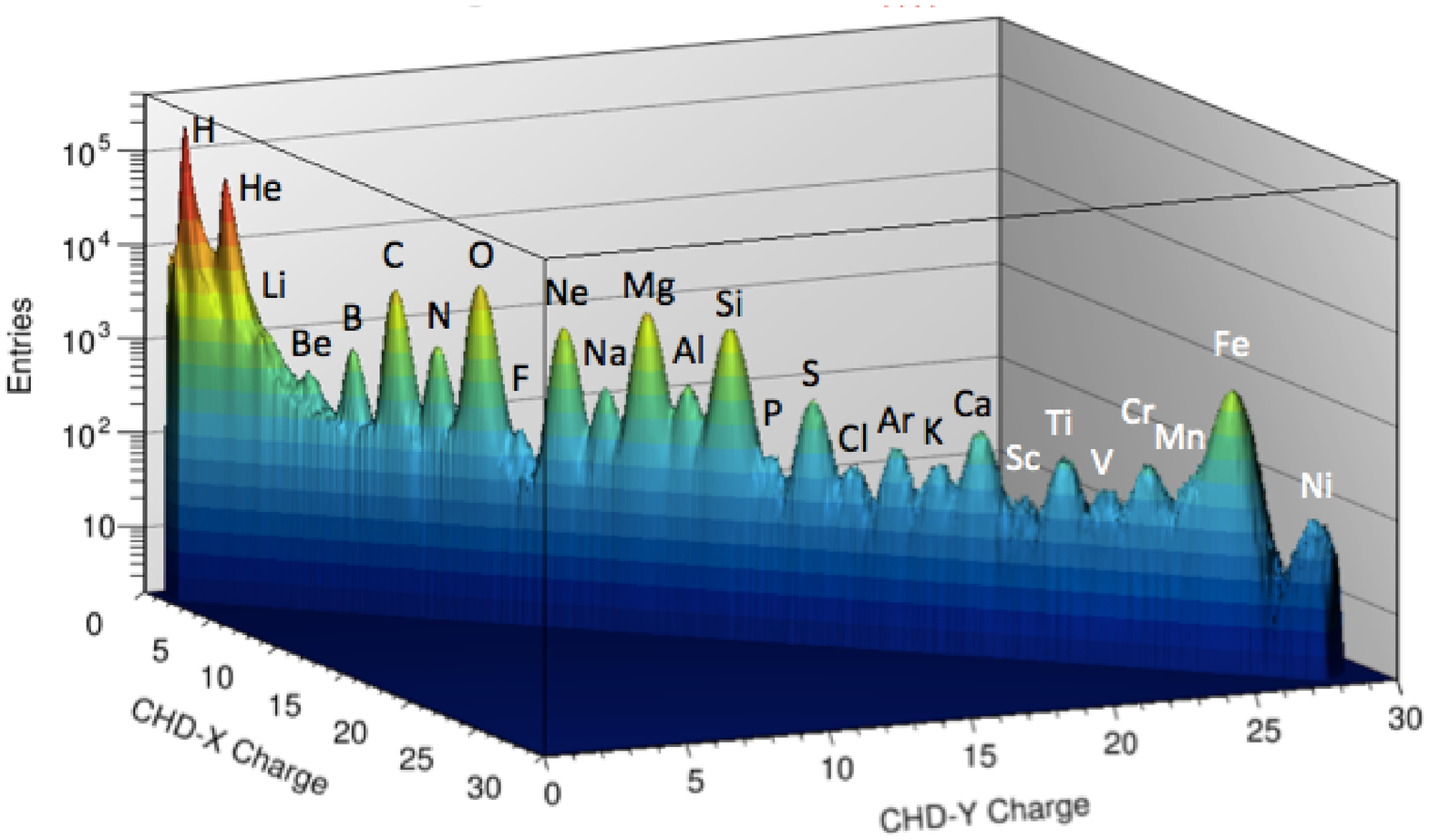}
    \end{center}
    \caption{\label{zdist2}Correlation between CHD-X and CHD-Y.}
  \end{minipage} 
\end{figure}

\paragraph{Detector calibration}
Energy calibration of each channel has been carried out by using penetrating cosmic-ray protons and helium~\cite{calib}.
The detector response includes position, temperature and time dependence of the plastic scintillators and PWO logs have been well studied and equalized with the MIP calibration.
For charge identification, the non-linearity of the CHD and IMC between detector response and deposit energy which is caused by the scintillation quenching effects is obtained from the flight data.
Figure~\ref{quench} shows the peak values of the CHD signal distributions of some charge derived from the flight data as a function of $Z^2$.
It is fitted by a function based on the halo model~\cite{quench} described as:
\begin{eqnarray}
f(Z^2)= A\left( \frac{1-f_h}{1+B_S(1-f_h)\alpha Z^2} + f_h \right) \alpha Z^2,
\label{eqCHD}
\end{eqnarray}
where $f_h$ is the ratio of halo region, $B_S$ is the quenching parameter, $A$ is the normalization factor and $\alpha$ is a constant (2~MeV).
Since the signal of CHD-X and Y are affected the $\delta$-ray from upstream material, the parameters of CHD-X and Y are obtained independently as shown in Fig.~\ref{quench}.
Figure~\ref{zdist2} shows the correlation of CHD-X and CHD-Y from proton to nickel.
Non-linearity of IMC has also been corrected with the same procedure used for the charge identification.

\paragraph{Track reconstruction}
The accurate reconstruction of shower axis and estimation of the impact point at the CHD are essential for the charge identification.
The shower axis is reconstructed by using the IMC signals.
Although heavy ions create many shower particles in the IMC, which could be a large background for the track reconstruction, 
the signals from the ionizing primary particle are commonly larger than the signals of the shower particles thanks to their large $dE/dx$ which is proportional to $Z^2$ disregarding the quenching effect.
The track is reconstructed by a least-square fit using the channels with maximum deposit in the upper four IMC layers.
The impact point at the CHD can be estimated with an accuracy of 330~$\mu$m for CHD-X and 300~$\mu$m for CHD-Y with minor charge and energy dependence.
Events whose reconstructed track is fully contained in the detector are selected for the nuclei analysis in this paper.

\begin{figure}[htbp]
  \begin{minipage}{17pc}
    \begin{center}
      \includegraphics[width=18pc]{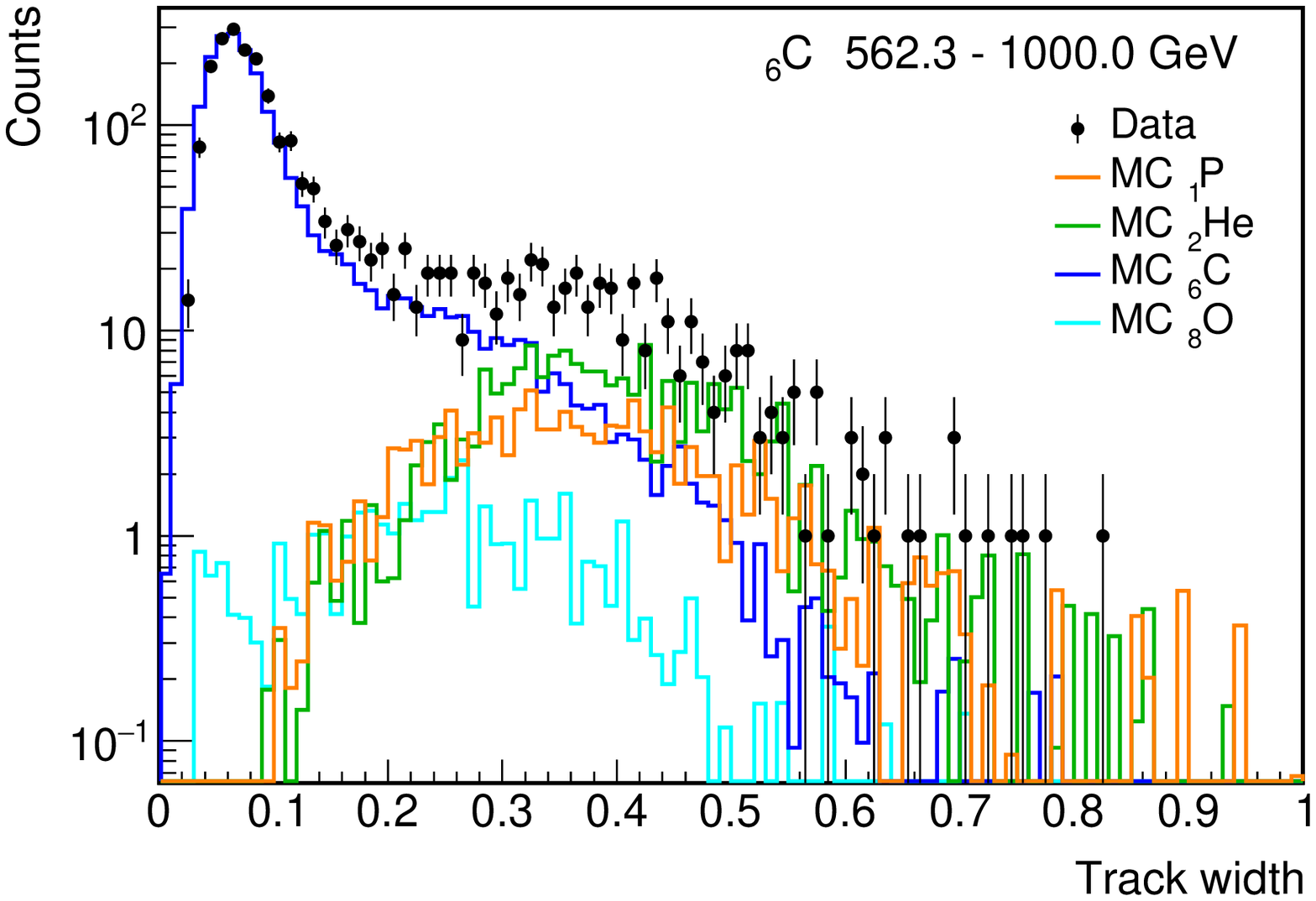}
    \end{center}
    \caption{\label{trkw} The distribution of track width for carbon before charge consistency cuts.}
  \end{minipage} 
  \hspace{2pc}%
  \begin{minipage}{17pc}
    \begin{center}
    \includegraphics[width=18pc]{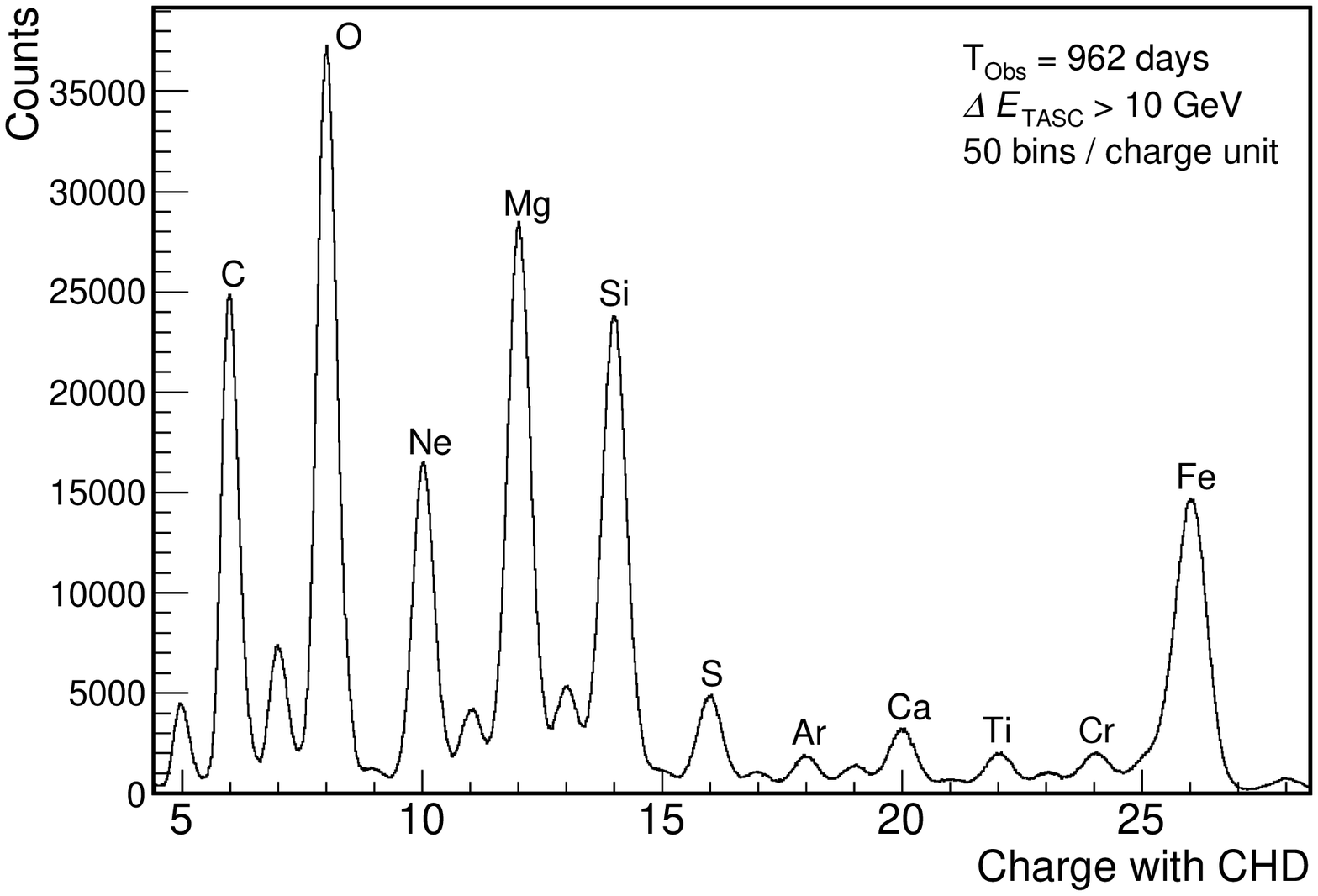}
    \end{center}
    \caption{\label{zdist}Charge distribution of CHD with $\Delta E_{\rm{TASC}}>10$~GeV.}
  \end{minipage}
\label{eventview}
\end{figure}

\begin{figure}[htbp]
  \begin{minipage}{17pc}
    \begin{center}
      \includegraphics[width=18pc]{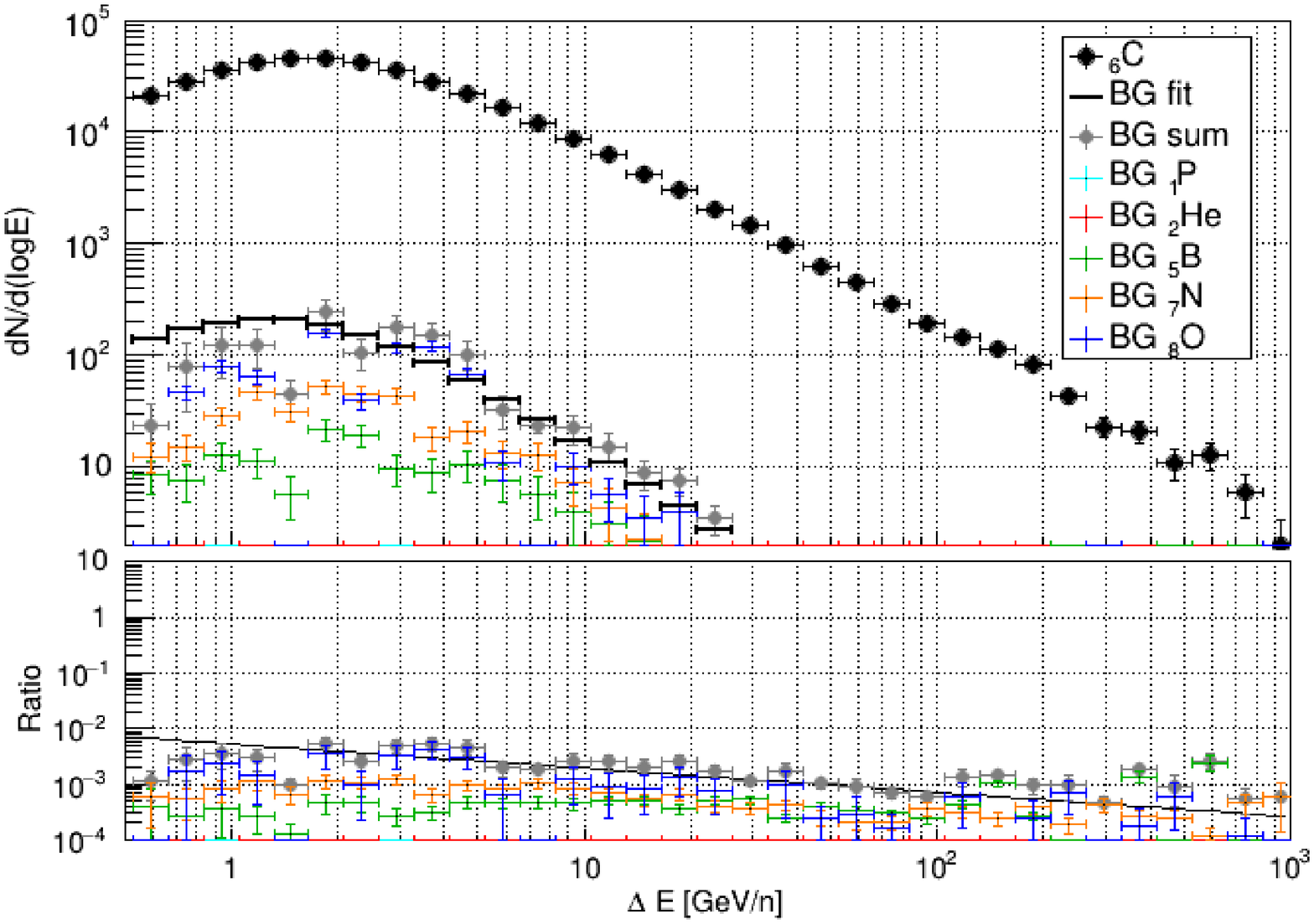}
    \end{center}
    \caption{\label{dNdE} dN/dE, estimated background and the ratio of background to signal for carbon.}
  \end{minipage}\hspace{2pc}%
  \begin{minipage}{17pc}
    \begin{center}
      \includegraphics[width=18pc]{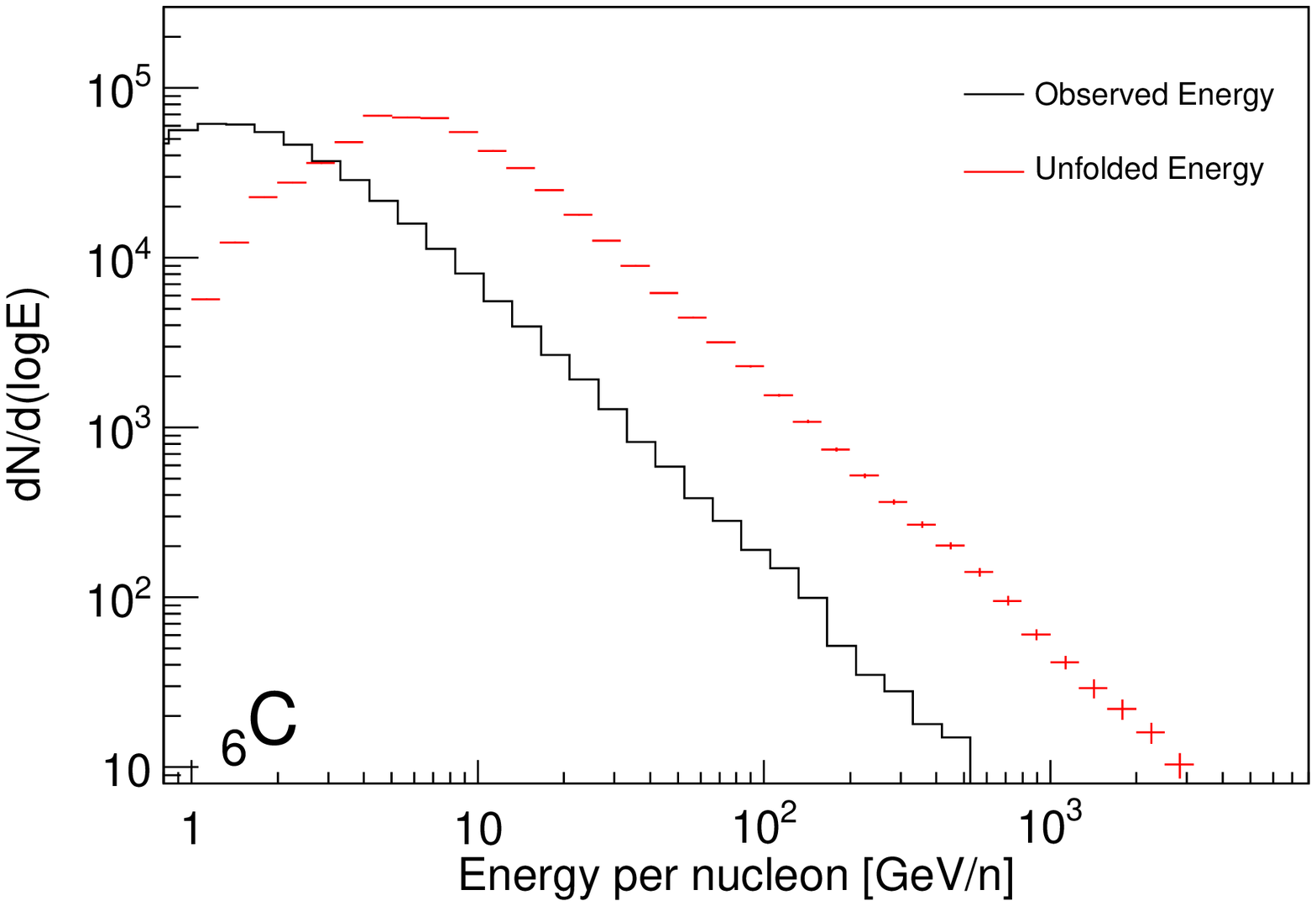}
    \end{center}
    \caption{\label{unfold} Energy distributions before and after the energy unfolding for carbon.}
  \end{minipage} 
\label{eventview}
\end{figure}

\paragraph{Charge identification}
To perform precise charge measurements and maintain good charge resolution, charge consistency of CHD-X, CHD-Y and upper IMC layers are required.
Events are selected if the difference of the charge with CHD-X and CHD-Y is less than 10\%.
For light nuclei with $Z\le8$, the charge consistency of IMC with 15\% is also required.
In addition, the track width defined as the difference of the sum of 7 SciFis and the sum of 3 SciFis associated with the reconstructed track and normalized to the reconstructed charge squared is used for the selection because the events interacting in the CHD make a wider track due to the secondary particles than that of penetrating events as shown in Fig.~\ref{trkw}.
Applying these cuts rejects events with poor charge resolution due to interactions in the CHD or upper IMC layers. 
The distribution of estimated charge in the range of $Z=5-28$ is shown in Fig.~\ref{zdist}.
Events within $\pm0.4$ charge units around the $Z$-peaks are identified as particular elements in the flux analysis presented in this paper.
The total efficiency is $\sim$30\% for carbon, $\sim$40\% for oxygen and more than $\sim$60\% for heavier particles.
The contamination from the other particles is estimated by means of MC simulations.
MC events in each energy bin for all species with $Z=1-28$ are normalized to the flight data based on the charge distribution with CHD. The number of contaminant events is calculated by the integration of all contaminant MC events within $\pm0.4$ charge unit.
Figure~\ref{dNdE} shows the energy distribution of carbon, the estimated background and the ratio.
The total background ratio is less than 1\% for carbon, oxygen and iron, less than 5\% for neon, magnesium and silicon.

\paragraph{Energy measurements and unfolding}
The shower energy for an event is determined by the sum of energy deposits in the TASC.
Since the energy leakage for nuclei from the calorimeter is unavoidable due to the characteristics of hadron induced showers and the finite detector resolution, the energy unfolding procedure for the derivation of the primary energy spectrum to correct bin-to-bin migration is necessary.
Iterative procedure based on the Bayes's theorem~\cite{bayes} with the RooUnfold package~\cite{unfold} is applied.
The response function for the energy unfolding is made from detailed MC simulations including the detector response.
MC events are selected with the same cuts used for flight data.
Figure~\ref{unfold} shows the distribution of observed energy and unfolded energy for carbon.

\section{Preliminary Results}
\begin{figure}[htbp]
  \begin{center}
    \includegraphics[width=31pc]{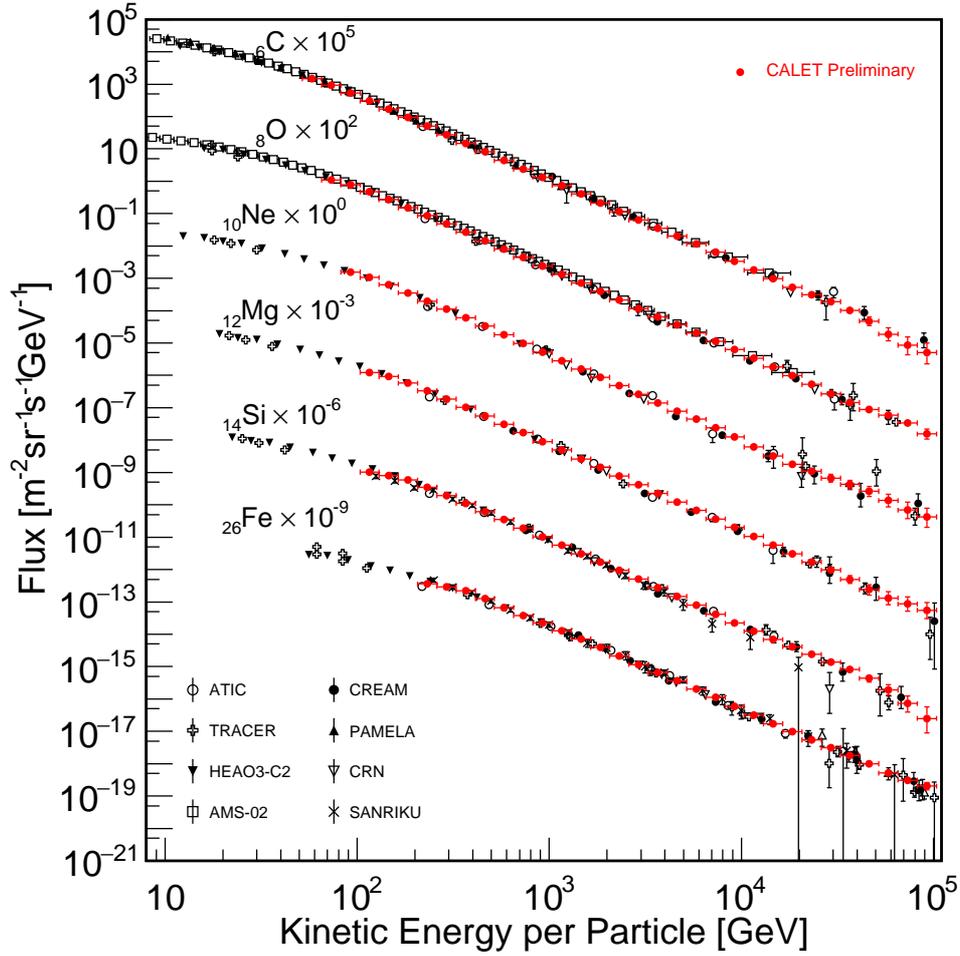}\hspace{2pc}%
    \begin{minipage}[b]{31pc}
      \caption{\label{flux} Preliminary energy spectra of carbon, oxygen, neon, magnesium, silicon and iron as a function of kinetic energy per particle after 962 days of CALET operation compared with previous observations~\cite{ams3,ams4,ATIC,CREAM,TRACER,PAMELA,HEAO3-C2,CRN,SANRIKU}. Only statistical errors are shown.}
    \end{minipage}
  \end{center}
\end{figure}

Energy flux, $\Phi(E)$, is calculated as follows,
\begin{eqnarray}
\Phi(E) = \frac{N(E)}{\varepsilon S \Omega T \Delta E}
\end{eqnarray}
where $N(E)$ is the number of events in the unfolded energy bin, $S\Omega$ is the geometrical acceptance, $T$ is live time, $\Delta E$ is the width of energy bin and $\varepsilon$ is the total efficiency.
Figure~\ref{flux} shows the preliminary energy spectra of carbon, oxygen, neon, magnesium, silicon and iron as a function of kinetic energy per particle with 962 days of operations.
These spectra are comparable with previous observations~\cite{ams3, ams4,ATIC,CREAM,TRACER,PAMELA,HEAO3-C2,CRN,SANRIKU}.
We note that the event selection we have applied here is based on a preliminary analysis as compared to what can eventually be achieved with CALET.
Further studies are now underway to provide a more detailed analysis of these spectra.

\section{Summary}
The ability of CALET to measure heavy cosmic-ray nuclei has been successfully demonstrated and preliminary energy spectra have been derived for the primary cosmic ray elements up to 100 TeV using 5.57$\times 10^6$ events in the range of C--Fe with $\Delta E>10$~GeV from the 962 days of operations.
The derived spectra illustrate the excellent capability of CALET to measure heavy ions with high statistics in a wide energy range.
Further studies on an increased data set and detailed systematic study will increase the sensitivity to detailed spectral features, which may be a key to solve questions about galactic cosmic-ray acceleration and propagation.

\medskip
\section*{References}
\bibliography{references}

\end{document}